\documentclass[a4paper,11pt]{article}
\usepackage{jheppub}
\usepackage{bm}
\usepackage{array}
\usepackage{booktabs}
\usepackage{multirow}
\usepackage{graphicx}
\usepackage{amsfonts}

\title{On Superconformal Index Degenerations and Half-Indices in S-Confining Gauge Theories}
\author[p]{Melisa Doğan} 
\author[p,q]{and Ilmar Gahramanov}

\affiliation[p]{Department of Physics, Bogazici University, 34342 Bebek, Istanbul, T\"urkiye}
\affiliation[q]{Center for Mathematics and its Applications, Khazar University, Mehseti St. 41, AZ1096, Baku, Azerbaijan}

\emailAdd{melisa.dogan@std.bogazici.edu.tr}
\emailAdd{ilmar.gahramanov@bogazici.edu.tr}

\abstract{We study the $p \rightarrow 0$ limit of four-dimensional $\mathcal N=1$ superconformal indices and its relation to half-indices of three-dimensional $\mathcal N=2$ theories. At the level of special functions, this limit reduces elliptic hypergeometric integrals to basic hypergeometric integrals. The scaling of the flavor fugacities determines the Neumann and Dirichlet boundary conditions of the chiral multiplets. The gauge multiplet has Neumann boundary conditions. We apply this prescription to s-confining theories with gauge groups $SU(2)$,
$USp(2N)$, $G_2$, and $SU(N)$, and recover the half-index identities of several boundary confining dualities, expressed through the Nassrallah-Rahman and Gustafson integrals. For $SU(N)$ with a rank-two antisymmetric chiral, four fundamentals and $N$ antifundamentals, we obtain an analytic proof of the half-index identities proposed by Okazaki and Smith for all $N \geq 5$.}

\begin{document}
\renewcommand{\afterProceedingsSpace}{\vskip21pt}
\renewcommand{\afterTitleSpace}{\vskip23pt}
\renewcommand{\afterRuleSpace}{\vskip23pt}
\renewcommand{\afterEmailSpace}{\vskip16pt}
\notoc
\vspace*{2cm}

\maketitle
\newpage

\section{Introduction}
The superconformal index is a useful tool for studying non-perturbative properties of supersymmetric gauge theories. In particular, indices of four-dimensional $\mathcal N=1$ theories can be written in terms of elliptic hypergeometric integrals \cite{Romelsberger2006,KinneyMaldacenaMinwallaRaju2007,DolanOsborn2009}. Equalities between the indices of dual theories then give identities for special functions \cite{SpiridonovVartanov2011,Rosengren:2017ujl}. This relation allows one to study supersymmetric dualities by analytic methods and, conversely, to interpret integral identities in terms of gauge theories.

An elementary example is the elliptic beta integral of Spiridonov \cite{Spiridonov2001},
\begin{equation}\label{eq:ebi}
\frac{(p;p)_\infty (q;q)_\infty}{2}
\int_{\mathbb{T}}
\frac{\displaystyle\prod_{i=1}^{6}
\Gamma(t_i z^{\pm1};p,q)}
{\Gamma(z^{\pm2};p,q)}
\frac{dz}{2\pi i z}
=
\prod_{1\le i<j\le6}
\Gamma(t_i t_j;p,q),
\qquad \prod_{i=1}^6 t_i = pq,
\end{equation}
This identity also expresses the equality of the superconformal indices of an s-confining $SU(2)$ theory and its confined description. Setting $t_6=pq/A$ with $A=\prod_{i=1}^{5}t_i$ and taking $p\to0$ gives the Nassrallah-Rahman integral \cite{NassrallahRahman1985,Rahman1986},
\begin{equation}\label{eq:NR}
\frac{(q;q)_\infty}{2}
\int_{\mathbb{T}}
\frac{
(A z^{\pm1};q)_\infty\,
(z^{\pm2};q)_\infty
}{
\displaystyle\prod_{i=1}^{5}
(t_i z^{\pm1};q)_\infty
}
\frac{dz}{2\pi i z}
=
\frac{
\displaystyle\prod_{j=1}^{5}
\left(A/t_j;q\right)_\infty
}{
\displaystyle\prod_{1\le i<j\le5}
(t_i t_j;q)_\infty
},
\qquad A\equiv\prod_{i=1}^{5}t_i .
\end{equation}
This provides an example of a basic hypergeometric integral identity obtained from an elliptic one.

Different limits of superconformal indices have different interpretations in supersymmetric field theory. For example, the hyperbolic limit of the four-dimensional index gives the partition function of a three-dimensional theory on $S^3$ \cite{DolanSpiridonovVartanov2011,GaddeYan2012,Imamura2011,Gahramanov:2013gpo,Gahramanov:2015tta}. Here we consider another limit, $p\to0$ with $q$ fixed. We identify its surviving contributions with those of the half-index of a three-dimensional $\mathcal N=2$ theory with boundary \cite{GaddeGukovPutrov2014,DimofteGaiottoPaquette2018}. The correspondence established here concerns protected indices and does not, by itself, imply a geometric dimensional reduction or an equivalence between the full theories.

The purpose of this paper is to derive this correspondence directly from the four-dimensional index and to illustrate it in several s-confining theories. The choice of scaling for the four-dimensional fugacities determines the boundary conditions of the corresponding three-dimensional multiplets. We first discuss this prescription in section~\ref{sec:general}. In section~\ref{sec:examples} we consider the elliptic beta integral and extend the analysis to $USp(2N)$, $G_2$, and $SU(N)$ gauge theories. In section~\ref{sec:antisym} we study the $SU(N)$ theory with an antisymmetric tensor and derive the half-index identities proposed in \cite{OkazakiSmith2023} for all $N\ge5$, including both even and odd $N$, by taking the $p\to0$ limit of known elliptic hypergeometric integral identities \cite{Spiridonov2003,Rains2010,SpiridonovVartanov2011}.

\section{From the 4d superconformal index to the 3d half-index}
\label{sec:general}

\subsection{The superconformal index and the $p\to0$ limit}

The four-dimensional $\mathcal N=1$ superconformal index \cite{Romelsberger2006,KinneyMaldacenaMinwallaRaju2007,DolanOsborn2009} counts BPS states on $S^3$ annihilated by a chosen supercharge and its conjugate,
\begin{equation}\label{eq:index}
I(p,q,u)=\mathrm{Tr}\,(-1)^F\,p^{\,j_1+j_2+\frac R2}\,q^{\,j_2-j_1+\frac R2}\,\prod_a u_a^{F_a},
\end{equation}
where $(j_1,j_2)$ denote the spins of $SU(2)_1\times SU(2)_2$, $R$ is an R-charge and $F_a$ are flavor charges. A chiral multiplet with R-charge $R$ and flavor fugacity $y$ contributes $\Gamma((pq)^{R/2}y;p,q)$. The parameters $p$ and $q$ are associated with the two rotations of $\mathbb C^2=\mathbb C_{w_1}\times\mathbb C_{w_2}$, together with the R-symmetry grading.

The dependence on the R-charges and flavor fugacities enters through $t=(pq)^{R/2}y$. Since an anomaly-free R-symmetry can mix with flavor symmetries, we choose a convenient R-charge $R_0$ for taking the limit. If $t=p^\alpha c$, with $c$ fixed, the corresponding field has $R_0=2\alpha$. The exponent of $p$ in the index is therefore
\begin{equation}
J_p\equiv j_1+j_2+\frac{R_0}{2}.
\end{equation}
At fixed $q$ and $c$, the limit selects states with $J_p=0$, provided that all BPS letters have non-negative $J_p$. For a chiral multiplet this requires $0\le R_0\le2$. Only the derivative carrying fugacity $q$ survives, so the remaining operators have a single holomorphic derivative. This agrees with the operator counting in the three-dimensional half-index on $\mathbb R_+\times\mathbb C$, where local boundary operators are graded by $q^{J+r/2}$ \cite{GaddeGukovPutrov2014,DimofteGaiottoPaquette2018}.

\subsection{Chiral multiplets and boundary conditions}

The single-letter index of a 4d chiral multiplet is
\begin{equation}\label{eq:chiral-letter}
i_{\chi}(t;p,q)=\frac{t-pq\,t^{-1}}{(1-p)(1-q)},\qquad t=(pq)^{R/2}y .
\end{equation}
The first term corresponds to the scalar $\phi$ and the second to the fermion $\bar\psi$. For $t=p^\alpha c$ they have $J_p=\alpha$ and $J_p=1-\alpha$, respectively. For $0\le\alpha\le1$ we obtain three possibilities:
\begin{itemize}
\item For $\alpha=0$ ($R_0=0$), the scalar letters survive, $i_\chi\to c/(1-q)$, giving the factor $1/(c;q)_\infty$ of a three-dimensional chiral multiplet with Neumann boundary condition \cite{GaddeGukovPutrov2014,DimofteGaiottoPaquette2018}.
\item For $\alpha=1$ ($R_0=2$), only the fermionic letters remain, $i_\chi\to-(q/c)/(1-q)$, giving the factor $(q/c;q)_\infty$ of a chiral multiplet with Dirichlet boundary condition. The usual three-dimensional parametrisation is $q/c=q^{1-r/2}y^{-1}$.
\item For $0<\alpha<1$, both terms vanish in the limit, and the multiplet gives a trivial contribution to the index.
\end{itemize}
These statements reproduce the limits of the elliptic gamma function (appendix~\ref{app:gamma})
\begin{equation}\label{eq:limits}
\lim_{p\to0}\Gamma(p^{\alpha}c;p,q)=
\begin{cases}
\dfrac{1}{(c;q)_\infty}, & \alpha=0 \quad\text{(Neumann)},\\[3mm]
1, & 0<\alpha<1 \quad\text{(decouples)},\\[1mm]
\left(q/c;q\right)_\infty, & \alpha=1 \quad\text{(Dirichlet)}.
\end{cases}
\end{equation}
With $c=q^{r/2}y$, where $r$ is the three-dimensional R-charge and $y$ a flavor fugacity, these are the half-index contributions $1/(q^{r/2}y;q)_\infty$ and $(q^{1-r/2}y^{-1};q)_\infty$ of Neumann and Dirichlet chiral multiplets. In practice a Dirichlet field appears with fugacity $pq/c$ and, by the reflection formula $\Gamma(x;p,q)\Gamma(pq/x;p,q)=1$, contributes $(c;q)_\infty$.

The same rule applies to gauge-invariant composite fields, with $R_0$ given by the sum of the charges of their constituents. In particular, the boundary conditions of mesons and baryons in the confined theory are determined by the same choice of $R_0$ as on the gauge theory side.

\subsection{Vector multiplets}

The vector multiplet has the single-letter index $\frac{2pq-p-q}{(1-p)(1-q)}\chi_{\mathrm{adj}}(z)$. Taking $p\to0$ gives $-\frac{q}{1-q}\chi_{\mathrm{adj}}(z)$. Including the Haar measure, we find
\begin{equation*}
(p;p)_\infty^{\mathrm{rk}}(q;q)_\infty^{\mathrm{rk}}\prod_{\alpha}\Gamma(z^\alpha;p,q)^{-1}\;\to\; (q;q)_\infty^{\mathrm{rk}}\prod_\alpha (z^\alpha;q)_\infty ,
\end{equation*}
which is the contribution of a three-dimensional $\mathcal N=2$ vector multiplet with Neumann boundary condition \cite{DimofteGaiottoPaquette2018}. Integration over the gauge fugacities projects onto gauge-invariant boundary operators.

\subsection{Summary and scope}

Under $p\to0$, a chiral multiplet with $R_0=0$ contributes with Neumann boundary condition, one with $R_0=2$ with Dirichlet boundary condition, and one with $0<R_0<2$ decouples; the vector multiplet contributes with Neumann boundary condition. When the limit exists and can be interchanged with the contour integral, an equality $I_A=I_B$ of four-dimensional indices yields an equality $\mathbb I_A=\mathbb I_B$ of half-indices.

Rescaling parameters by powers of $p$ and using the reflection formula is a standard way of taking such limits \cite{Spiridonov2008,SpiridonovVartanov2011,RainsLimits2009}. Here we identify the scaling exponents with boundary conditions. The construction concerns protected operator counts. By itself, it does not provide a dimensional reduction obtained by shrinking a circle or establish an equivalence between the full three-dimensional theories.

\section{Boundary confining dualities from known $q$-beta integrals}
\label{sec:examples}
\subsection{$SU(2)$ with six doublets}


We start with the s-confining $SU(2)$ gauge theory with six doublets \cite{Seiberg1994a,SpiridonovVartanov2011,Gahramanov:2013xsa}. Its dual description consists of 15 mesons $M=Q^2$ in the antisymmetric representation of $SU(6)$ with superpotential $W=\mathrm{Pf}\,M$. The equality of their indices is the elliptic beta integral~\eqref{eq:ebi}.

We keep $t_1,\ldots,t_5$ fixed and set $t_6=pq/A$ with $A=\prod_{i=1}^5t_i$. By the reflection formula,
\begin{equation}
\Gamma(t_{6}z^{\pm1};p,q) = \frac{1}{\Gamma(A z^{\mp1};p,q)},\qquad
\Gamma (t_{m}t_{6};p,q) = \frac{1}{\Gamma (A/t_m;p,q)},
\end{equation}
so that every elliptic gamma function has a $p$-independent argument. Taking $p\to0$ using eq.~\eqref{eq:limits} gives the Nassrallah-Rahman integral~\eqref{eq:NR}. The five fixed doublets have $R_0=0$ and Neumann boundary conditions, and the sixth doublet $\tilde Q$ has $R_0=2$ and Dirichlet boundary condition; the factor $(Az^{\pm1};q)_\infty$ in~\eqref{eq:NR} is its contribution. The boundary anomaly is $2-\tfrac52+\tfrac12=0$.

To compare with \cite{OkazakiSmith2023}, we split the five Neumann doublets as $5=3+2$ and set
\begin{equation}\label{eq:NRfug}
t_I = q^{r_a/2}a\,x_I \ (I=1,2,3),\qquad t_{3+\alpha} = q^{r_b/2}b\,\tilde x_\alpha \ (\alpha=1,2),\qquad
\prod_{I=1}^{3}x_I=\prod_{\alpha=1}^{2}\tilde{x}_\alpha=1,
\end{equation}
so that $A=q^{\frac{3r_a+2r_b}{2}}a^3b^2$. The half-indices of the two theories are
\begin{equation}
\mathbb I_A=\frac{(q;q)_\infty}{2} \int_{\mathbb{T}}  \frac{dz}{2{\pi}iz} \frac{(z^{\pm2};q)_\infty\, (q^{\frac{3r_a+2r_b}{2}}a^3b^2z^{\pm1};q)_\infty} {\prod_{I=1}^{3}(q^{\frac{r_a}{2}}ax_{I}z^{\pm1};q)_\infty \prod_{\alpha=1}^{2}(q^{\frac{r_b}{2}}b\tilde{x}_{\alpha}z^{\pm1};q)_\infty}
\end{equation}
and
\begin{equation}
\mathbb I_B=\frac{\prod_{I=1}^{3} (q^{r_a+r_b}a^2b^2 x_{I}^{-1};q)_\infty  \prod_{\alpha=1}^{2} (q^{\frac{3r_a+r_b}{2}}a^3b\, \tilde{x}_{\alpha}^{-1};q)_\infty}{\prod_{I=1}^{3}(q^{r_a}a^2x_{I}^{-1};q)_\infty\, (q^{r_{b}}b^2;q)_\infty   \prod_{I=1}^{3} \prod_{\alpha=1}^{2}   (q^{\frac{r_a+r_b}{2}}ab\,x_{I}\tilde{x}_{\alpha};q)_\infty} ,
\end{equation}
The equality $\mathbb I_A=\mathbb I_B$ is the half-index identity of the boundary confining duality of \cite{OkazakiSmith2023}. The field content is given in Table~\ref{tab:NR}.

\begin{table}[ht]
\centering
\small
\renewcommand{\arraystretch}{1.3}
\begin{tabular}{|c|c|c|c|c|}
\hline
 & $SU(2)$ & $SU(3)\times SU(2)$ & fugacity & b.c.\\
\hline
\multicolumn{5}{|l|}{\textbf{Theory A}}\\
\hline
VM & adj & $(1,1)$ & - & $\mathcal N$\\
$Q_I$ & $\square$ & $(\square,1)$ & $t_I=q^{r_a/2}a\,x_I$ & $\mathcal N$\\
$Q_\alpha$ & $\square$ & $(1,\square)$ & $t_{3+\alpha}=q^{r_b/2}b\,\tilde x_\alpha$ & $\mathcal N$\\
$\tilde Q$ & $\square$ & $(1,1)$ & $t_6=pq/A$ & $\mathcal D$\\
\hline
\multicolumn{5}{|l|}{\textbf{Theory B}}\\
\hline
$M_{I\alpha}\sim Q_IQ_\alpha$ & $1$ & $(\square,\square)$ & $t_It_{3+\alpha}$ & $\mathcal N$\\
$B_I\sim \epsilon_{IJK}Q_JQ_K$ & $1$ & $(\overline{\square},1)$ & $t_Jt_K$ & $\mathcal N$\\
$B\sim \epsilon_{\alpha\beta}Q_\alpha Q_\beta$ & $1$ & $(1,1)$ & $t_4t_5$ & $\mathcal N$\\
$\tilde B_\alpha\sim Q_\alpha\tilde Q$ & $1$ & $(1,\square)$ & $pq/(A/t_{3+\alpha})$ & $\mathcal D$\\
$M^f_I\sim Q_I\tilde Q$ & $1$ & $(\square,1)$ & $pq/(A/t_I)$ & $\mathcal D$\\
\hline
\end{tabular}
\caption{Dual pair for the Nassrallah-Rahman limit of the $SU(2)$ theory with six doublets, $A=\prod_{i=1}^5t_i$. In the row of $B_I$, $\{I,J,K\}=\{1,2,3\}$.}
\label{tab:NR}
\end{table}

\subsection{$USp(2N)$ with $2N+4$ fundamentals}
\label{sec:USp}

The $USp(2N)$ theory with $2N+4$ fundamentals is s-confining \cite{IntriligatorPouliot1995}. Its dual description consists of mesons $M=Q^2$ in the $\Lambda^2$ of $SU(2N+4)$. The equality of the indices is the type I $BC_N$ elliptic beta integral \cite{vanDiejenSpiridonov2000,Rains2010},
\begin{multline}
\frac{(p;p)^N_\infty (q;q)^N_\infty}{2^N N!} \int_{\mathbb{T}^N} \prod_{1 \leq i < j \leq N} \frac{1}{\Gamma(z_{i}^{\pm1} z_{j}^{\pm1};p,q) }
\times\prod_{j=1}^{N} \frac{\prod_{m=1}^{2N+4} \Gamma(t_m z_{j}^{\pm1};p,q)}{\Gamma (z_{j}^{\pm2};p,q) } \frac{dz_j}{2{\pi}iz_j}\\
=\prod_{1 \leq m < j \leq 2N+4} \Gamma(t_mt_j;p,q),
\end{multline}
with $\prod_{m=1}^{2N+4}t_m = pq$. Setting $t_{2N+4}=pq/A$ with $A\equiv\prod_{m=1}^{2N+3}t_m$, applying reflection to the factors containing $t_{2N+4}$ and taking $p\to0$ gives
\begin{multline}\label{eq:USpA}
\frac{(q;q)^N_\infty}{2^N N!} \int_{\mathbb{T}^N} \prod_{1 \leq i < j \leq N} ( z_i^{\pm1}z_j^{\pm1};q)_\infty
\times\prod_{j=1}^{N} \frac{(z_j^{\pm2};q)_\infty\,(A\,z_j^{\pm1};q)_\infty}{\prod_{m=1}^{2N+3} (t_mz_j^{\pm1};q)_\infty} \frac{dz_j}{2{\pi}iz_j}\\
=\frac{\prod_{j=1}^{2N+3} (A/t_j;q)_\infty}{\prod_{1 \leq m < j \leq 2N+3} (t_mt_j;q)_\infty},
\end{multline}
This is Gustafson's multivariable generalisation of the Nassrallah-Rahman integral \cite{Gustafson1992SIAM,Gustafson1994b}. For $N=1$ it reduces to~\eqref{eq:NR}. With $t_m=q^{r/2}a\,x_m$ and
$\prod_{m=1}^{2N+3}x_m=1$, we have
$A=q^{\frac{(2N+3)r}{2}}a^{2N+3}$.
The half-index identity becomes
\begin{multline}
\frac{(q;q)^N_\infty}{2^N N!}
\int_{\mathbb{T}^N}
\prod_{1\leq i<j\leq N}
(z_i^{\pm1}z_j^{\pm1};q)_\infty 
\times\prod_{j=1}^{N}
\frac{
(z_j^{\pm2};q)_\infty
(q^{\frac{(2N+3)r}{2}}a^{2N+3}z_j^{\pm1};q)_\infty
}{
\prod_{m=1}^{2N+3}
(q^{r/2}ax_mz_j^{\pm1};q)_\infty
}
\frac{dz_j}{2\pi i z_j}
\\
=
\frac{
\prod_{j=1}^{2N+3}
(q^{\frac{(2N+2)r}{2}}a^{2N+2}x_j^{-1};q)_\infty
}{
\prod_{1\leq m<j\leq 2N+3}
(q^ra^2x_mx_j;q)_\infty
}.
\end{multline}
This is the half-index identity of the boundary
confining duality of
\cite{OkazakiSmith2023},
with field content given in Table~\ref{tab:USp}.

\begin{table}[ht]
\centering
\small
\renewcommand{\arraystretch}{1.3}
\begin{tabular}{|c|c|c|c|c|}
\hline
 & $USp(2N)$ & $SU(2N+3)$ & fugacity & b.c.\\
\hline
\multicolumn{5}{|l|}{\textbf{Theory A}}\\
\hline
VM & adj & $1$ & - & $\mathcal N$\\
$Q_\alpha$ & $\square$ & $\square$ & $t_\alpha=q^{r/2}a\,x_\alpha$ & $\mathcal N$\\
$Q_{2N+4}$ & $\square$ & $1$ & $pq/A$ & $\mathcal D$\\
\hline
\multicolumn{5}{|l|}{\textbf{Theory B}}\\
\hline
$M_{\alpha\beta}\sim Q_\alpha Q_\beta$ & $1$ & $\Lambda^2$ & $t_\alpha t_\beta$ & $\mathcal N$\\
$M_{\alpha f}\sim Q_\alpha Q_{2N+4}$ & $1$ & $\square$ & $pq/(A/t_\alpha)$ & $\mathcal D$\\
\hline
\end{tabular}
\caption{Dual pair for the $USp(2N)$ theory with $2N+4$ fundamentals, $\alpha=1,\ldots,2N+3$ and $A=\prod_{\alpha=1}^{2N+3}t_\alpha$.}
\label{tab:USp}
\end{table}

\subsection{$G_2$ with five fundamentals}
\label{sec:G2}

The $G_2$ theory with five fundamentals is s-confining \cite{CsakiSchmaltzSkiba1997}. Its confined description is given in terms of $Q^2$, $Q^3$ and $Q^4$. Its index identity \cite{SpiridonovVartanov2011} reads
\begin{multline}
\frac{(p;p)^2_\infty (q;q)^2_\infty} {2^2\cdot 3} \prod_{m=1}^{5} \Gamma(t_m;p,q) \int_{\mathbb{T}^2} \frac{\prod_{k=1}^{3} \prod_{m=1}^{5}\Gamma(t_m z_k^{\pm1};p,q)}{\prod_{1 \leq j < k \leq 3}\Gamma (z_j^{\pm1}z_k^{\pm1};p,q)} \prod_{j=1}^{2}\frac{dz_j}{2{\pi}iz_j}\\
=\prod_{m=1}^{5} \frac{\Gamma( t_m^2;p,q)
}{\Gamma((pq)^{\frac{1}{2}}t_m;p,q)} \prod_{1 \leq l < m \leq 5} \frac{\Gamma( t_lt_m;p,q)
}{\Gamma((pq)^{\frac{1}{2}}t_lt_m;p,q)},
\end{multline}
with $z_1z_2z_3=1$ and $\prod_{m=1}^{5} t_m = (pq)^{1/2}$; the denominators on the right are the contributions of $Q^4$ and $Q^3$, rewritten with the reflection formula and the balancing condition. As noted in \cite{SpiridonovVartanov2011}, substituting $t_5=(pq)^{1/2}/A_4$ with $A_4\equiv\prod_{m=1}^4t_m$ and taking $p\to0$ gives Gustafson's $G_2$ integral \cite{Gustafson1994}. Here we recover this half-index identity from the $p\to0$ limit of the four-dimensional superconformal index.

The arguments $t_5$, $t_mt_5$, $(pq)^{1/2}t_m$ and $(pq)^{1/2}t_lt_m$ scale as $p^{1/2}$, so the fifth fundamental and the composites containing it once decouple. The remaining factors follow from $\Gamma(t_5^2)=1/\Gamma(A_4^2)$, $\Gamma((pq)^{\frac12}t_5)=1/\Gamma(A_4)$ and $\Gamma((pq)^{\frac12}t_mt_5)=1/\Gamma(A_4/t_m)$. With $t_\alpha=q^{r/2}ax_\alpha$, $\prod_\alpha x_\alpha=1$, and
$\prod_{1 \leq j < k \leq 3} (z_j^{\pm1}z_k^{\pm1};q)_\infty=\prod_{j\neq k}(z_jz_k^{-1};q)_\infty\prod_{i=1}^{3}(z_i^{\pm1};q)_\infty$, the limit reads
\begin{equation}\label{eq:G2half}
\begin{aligned}
&\frac{(q;q)_\infty^2}{2^2\cdot 3}
\int_{\mathbb{T}^2}
\frac{
\displaystyle
\prod_{\substack{1\le j,k\le3\\j\neq k}}
(z_j z_k^{-1};q)_\infty
\prod_{i=1}^{3}(z_i^{\pm1};q)_\infty
}{
\displaystyle
\prod_{\alpha=1}^{4}\Big[
(q^{\frac r2}a x_\alpha;q)_\infty
\prod_{i=1}^{3}
(q^{\frac r2}a x_\alpha z_i^{\pm1};q)_\infty\Big]
}
\prod_{i=1}^{2}
\frac{dz_i}{2\pi i z_i}
\\[2mm]
&\hspace{2cm}
=
\frac{(q^{4r}a^8;q)_\infty}
{
(q^{2r}a^4;q)_\infty
\displaystyle\prod_{1\le\alpha\le\beta\le4}
(q^r a^2 x_\alpha x_\beta;q)_\infty
\displaystyle\prod_{\alpha=1}^{4}
(q^{\frac{3r}{2}}a^3 x_\alpha^{-1};q)_\infty
},
\end{aligned}
\end{equation}
which reproduces the half-index identity of the $G_2$ boundary confining duality studied by Okazaki and Smith \cite{OkazakiSmithExceptional2023}. The four remaining fundamentals and the vector multiplet are Neumann, with boundary anomaly $4-4=0$ ($T(\mathbf 7)=1$). On the confined side, the Dirichlet factor associated with $V$ is reproduced by the $p\to0$ limit of the contribution from the four-dimensional composite $Q_5Q_5$. This correspondence holds at the level of the index and does not imply an identification of the operators. The Neumann singlet $B$ is matched by $Q_1Q_2Q_3Q_4$, and $B_\alpha$ by the $Q^3$ composites constructed from the first four flavors. The terms with $\alpha=\beta$ in~\eqref{eq:G2half} arise from $(t_\alpha^2;q)_\infty$. See Table~\ref{tab:G2}.

\begin{table}[ht]
\centering
\small
\renewcommand{\arraystretch}{1.3}
\begin{tabular}{|c|c|c|c|c|}
\hline
 & $G_2$ & $SU(4)$ & fugacity & b.c.\\
\hline
\multicolumn{5}{|l|}{\textbf{Theory A}}\\
\hline
VM & adj & $1$ & - & $\mathcal N$\\
$Q_\alpha$ & $\mathbf 7$ & $\square$ & $t_\alpha=q^{r/2}a\,x_\alpha$ & $\mathcal N$\\
$Q_5$ & $\mathbf 7$ & $1$ & $(pq)^{1/2}/A_4$ & -\\
\hline
\multicolumn{5}{|l|}{\textbf{Theory B}}\\
\hline
$M_{\alpha\beta}\sim Q_\alpha Q_\beta$ & $1$ & $\mathrm{Sym}^2$ & $t_\alpha t_\beta$ & $\mathcal N$\\
$B_\alpha\sim \epsilon_{\alpha\beta\gamma\delta}Q_\beta Q_\gamma Q_\delta$ & $1$ & $\overline{\square}$ & $A_4/t_\alpha$ & $\mathcal N$\\
$B\sim Q_1Q_2Q_3Q_4$ & $1$ & $1$ & $A_4$ & $\mathcal N$\\
$V$ & $1$ & $1$ & $pq/A_4^2$ & $\mathcal D$\\
$Q_\alpha Q_5$, $Q^3$ and $Q^4$ containing $Q_5$ & $1$ & various & $\propto p^{1/2}$ & -\\
\hline
\end{tabular}
\caption{Dual pair for the $G_2$ theory with five fundamentals, $\alpha=1,\ldots,4$ and $A_4=\prod_{\alpha=1}^4t_\alpha$. The fifth fundamental and all composites containing it exactly once have $R_0=1$ and decouple.}
\label{tab:G2}
\end{table}

\subsection{$SU(N)$ with $N+1$ flavors}
\label{sec:SUN}

The $SU(N)$ theory with $N+1$ flavors is s-confining \cite{Seiberg1994a}. The confined description contains mesons $Q\tilde Q$, baryons $Q^N$ and antibaryons $\tilde Q^N$. The equality of the indices is the $A_{N-1}$ elliptic beta integral, conjectured in \cite{Spiridonov2003} and proven in \cite{Rains2010} (see also \cite{SpiridonovVartanov2011}):
\begin{multline}
\frac{(p;p)_\infty^{N-1} (q;q)_\infty^{N-1}}{N!} \int_{\mathbb{T}^{N-1}}   \prod_{1 \leq i < j \leq N } \frac{1}{\Gamma(z_iz_j^{-1},z_i^{-1}z_j ; p,q)}
\times\prod_{j = 1 }^ {N}  \prod_{m = 1 }^ {N+1} \Gamma ( s_m z_j,t_mz_j^{-1};p,q) \prod_{j = 1 }^ {N-1} \frac{dz_j}{2\pi iz_j}\\
= \prod_{m = 1 }^ {N+1} \Gamma(Ss_m^{-1} , Tt_m^{-1};p,q) \prod_{k,m = 1 }^ {N+1} \Gamma(s_k t_m;p,q),
\end{multline}
with $\prod_jz_j=1$, $S = \prod_{m} s_m$, $T = \prod_{m} t_m$ and $ST = pq$. We set $t_{N+1}=pq/(ST')$ with $T'\equiv\prod_{m=1}^{N}t_m$. Then $\Gamma(t_{N+1}z_j^{-1})=1/\Gamma(ST'z_j)$, $\Gamma(T/t_m)=1/\Gamma(St_m)$ for $m\le N$, $\Gamma(T/t_{N+1})=\Gamma(T')$ and $\Gamma(s_kt_{N+1})=1/\Gamma(ST'/s_k)$, and the limit $p\to0$ gives
\begin{multline}\label{eq:SUN}
\frac{(q;q)_\infty^{N-1}}{N!}\int_{\mathbb{T}^{N-1}}\prod_{i\neq j}(z_iz_j^{-1};q)_\infty
\times\prod_{j=1}^{N}\frac{(ST'z_j;q)_\infty}{\prod_{m=1}^{N+1}(s_mz_j;q)_\infty\prod_{m=1}^{N}(t_mz_j^{-1};q)_\infty}
\prod_{j=1}^{N-1}\frac{dz_j}{2\pi iz_j}\\
=\frac{\prod_{m=1}^{N}(St_m;q)_\infty\prod_{k=1}^{N+1}(ST'/s_k;q)_\infty}
{(T';q)_\infty\prod_{m=1}^{N+1}(S/s_m;q)_\infty\prod_{k=1}^{N+1}\prod_{m=1}^{N}(s_kt_m;q)_\infty} .
\end{multline}
This is Gustafson's $A_{N-1}$ generalisation of the Nassrallah-Rahman integral \cite{Gustafson1994b}; that the $p\to0$ limit of this elliptic identity reduces to a Gustafson integral is also noted in \cite{SpiridonovVartanov2011}.

On the gauge side, $\tilde Q_{N+1}$ is Dirichlet and all other chirals are Neumann, with boundary anomaly $N-\tfrac{2N+1}{2}+\tfrac12=0$. With $s_k=q^{r_a/2}a\,x_k$ and
$t_m=q^{r_b/2}b\,\tilde{x}_m$ for $m\leq N$,
where $\prod_{k=1}^{N+1}x_k=1$ and
$\prod_{m=1}^{N}\tilde{x}_m=1$, we have
$S=q^{\frac{(N+1)r_a}{2}}a^{N+1}$ and
$T'=q^{\frac{Nr_b}{2}}b^N$.
Substituting into \eqref{eq:SUN}, the half-index
identity becomes
\begin{multline}
\frac{(q;q)_\infty^{N-1}}{N!}
\int_{\mathbb{T}^{N-1}}
\prod_{i\neq j}(z_iz_j^{-1};q)_\infty \prod_{j=1}^{N}
\frac{
(q^{\frac{(N+1)r_a+Nr_b}{2}}
a^{N+1}b^Nz_j;q)_\infty
}{
\prod_{k=1}^{N+1}
(q^{r_a/2}ax_kz_j;q)_\infty
\prod_{m=1}^{N}
(q^{r_b/2}b\tilde{x}_mz_j^{-1};q)_\infty
}
\prod_{j=1}^{N-1}\frac{dz_j}{2\pi iz_j}
\\
=
\frac{
\prod_{k=1}^{N+1}
(q^{\frac{N(r_a+r_b)}{2}}
a^Nb^Nx_k^{-1};q)_\infty
\prod_{m=1}^{N}
(q^{\frac{(N+1)r_a+r_b}{2}}
a^{N+1}b\tilde{x}_m;q)_\infty
}{
\prod_{k=1}^{N+1}
(q^{\frac{Nr_a}{2}}a^Nx_k^{-1};q)_\infty
(q^{\frac{Nr_b}{2}}b^N;q)_\infty
\prod_{k=1}^{N+1}\prod_{m=1}^{N}
(q^{\frac{r_a+r_b}{2}}
abx_k\tilde{x}_m;q)_\infty
}.
\end{multline}
This is the half-index identity of the boundary
confining duality of
\cite{OkazakiSmith2023},
with field content given in Table~\ref{tab:SUN}: $M_{km}$, $B_k$ and $\tilde B_{N+1}$ correspond to their Neumann fields $M_{I\alpha}$, $B^I$ and $B$, while $\tilde B_m$ and $M_{k,N+1}$ correspond to their Dirichlet fields $\tilde B_\alpha$ and $\tilde M_I$.

\begin{table}[ht]
\centering
\small
\renewcommand{\arraystretch}{1.3}
\begin{tabular}{|c|c|c|c|c|}
\hline
 & $SU(N)$ & $SU(N+1)\times SU(N)$ & fugacity & b.c.\\
\hline
\multicolumn{5}{|l|}{\textbf{Theory A}}\\
\hline
VM & adj & $(1,1)$ & - & $\mathcal N$\\
$Q_k$ & $\square$ & $(\square,1)$ & $s_k=q^{r_a/2}a\,x_k$ & $\mathcal N$\\
$\tilde Q_m$ & $\overline{\square}$ & $(1,\square)$ & $t_m=q^{r_b/2}b\,\tilde x_m$ & $\mathcal N$\\
$\tilde Q_{N+1}$ & $\overline{\square}$ & $(1,1)$ & $pq/(ST')$ & $\mathcal D$\\
\hline
\multicolumn{5}{|l|}{\textbf{Theory B}}\\
\hline
$M_{km}\sim Q_k\tilde Q_m$ & $1$ & $(\square,\square)$ & $s_kt_m$ & $\mathcal N$\\
$B_k\sim \epsilon\, Q^{N}$ & $1$ & $(\overline{\square},1)$ & $S/s_k$ & $\mathcal N$\\
$\tilde B_{N+1}\sim \epsilon\,\tilde Q_1\cdots\tilde Q_N$ & $1$ & $(1,1)$ & $T'$ & $\mathcal N$\\
$M_{k,N+1}\sim Q_k\tilde Q_{N+1}$ & $1$ & $(\square,1)$ & $pq/(ST'/s_k)$ & $\mathcal D$\\
$\tilde B_m\sim \epsilon\,\tilde Q^{N-1}\tilde Q_{N+1}$ & $1$ & $(1,\overline{\square})$ & $pq/(St_m)$ & $\mathcal D$\\
\hline
\end{tabular}
\caption{Dual pair for the $SU(N)$ theory with $N+1$ flavors, $k=1,\ldots,N+1$, $m=1,\ldots,N$, $S=\prod_{k=1}^{N+1}s_k$ and $T'=\prod_{m=1}^{N}t_m$. Here $B_k$ omits $Q_k$ and $\tilde B_m$ omits $\tilde Q_m$.}
\label{tab:SUN}
\end{table}


\section{New $q$-beta integrals from $SU(N)$ with an antisymmetric tensor}
\label{sec:antisym}
We now consider four-dimensional $SU(N)$ gauge theory with one rank-two antisymmetric chiral multiplet, four fundamentals and $N$ antifundamentals, which is s-confining \cite{CsakiSchmaltzSkiba1997,CsakiSchmaltzSkiba1997PRD}. The equality of the superconformal indices of the electric and confined descriptions is given\footnote{See also \cite{Rosengren:2017ujl} for a survey of related elliptic hypergeometric integral evaluations associated with root systems.} in \cite{SpiridonovVartanov2011} (see also \cite{Spiridonov2003,Rains2010}).  We use these established identities to derive the half-index relations of \cite{OkazakiSmith2023}.

\subsection{Four-dimensional index identities}

Denote the antisymmetric fugacity by $a$, the fundamental fugacities by $s_I$ ($I=1,\ldots,4$), and the antifundamental fugacities by $t_j$ ($j=1,\ldots,N$). Define
\begin{equation}
 S=\prod_{I=1}^{4}s_I,\qquad T=\prod_{j=1}^{N}t_j,
 \qquad a^{N-2}ST=pq.
 \label{eq:odd-antisymmetric-balancing}
\end{equation}
With $\prod_{i=1}^{N}z_i=1$, the electric index reads
\begin{equation}
\begin{aligned}
 I_E={}&\frac{(p;p)_\infty^{N-1}(q;q)_\infty^{N-1}}{N!}
 \int_{\mathbb T^{N-1}}\prod_{i=1}^{N-1}\frac{dz_i}{2\pi i z_i}
 \frac{\prod_{1\leq i<j\leq N}\Gamma(az_iz_j;p,q)}
 {\prod_{\substack{1\leq i,j\leq N\\i\ne j}}\Gamma(z_i/z_j;p,q)}
 \\
 &\times\prod_{i=1}^{N}\prod_{I=1}^{4}\Gamma(s_Iz_i;p,q)
 \prod_{i=1}^{N}\prod_{j=1}^{N}\Gamma(t_j/z_i;p,q).
\end{aligned}
\label{eq:odd-antisymmetric-4d-electric}
\end{equation}
For $N=2n+1$ the confined index is
\begin{equation}
\begin{aligned}
 I_C^{(2n+1)}={}&\prod_{1\leq j<k\leq N}\Gamma(at_jt_k;p,q)
 \prod_{I=1}^{4}\prod_{j=1}^{N}\Gamma(s_It_j;p,q)
 \\
 &\times\Gamma(T;p,q)
 \prod_{I=1}^{4}\frac{\Gamma(a^n s_I;p,q)}
 {\Gamma(a^nTs_I;p,q)},
\end{aligned}
\label{eq:odd-antisymmetric-4d-confined}
\end{equation}
and for $N=2n$ it is
\begin{equation}
\begin{aligned}
 I_C^{(2n)}={}&\prod_{1\leq j<k\leq N}\Gamma(at_jt_k;p,q)
 \prod_{I=1}^{4}\prod_{j=1}^{N}\Gamma(s_It_j;p,q)
 \\
 &\times\frac{\Gamma(a^n,T;p,q)}{\Gamma(a^nT;p,q)}
 \prod_{1\le I<J\le4}\Gamma(a^{n-1}s_Is_J;p,q).
\end{aligned}
\label{eq:even-antisymmetric-4d-confined}
\end{equation}
In both cases $I_E=I_C$. The parameter identification with Theorem~7 of \cite{Spiridonov2003} is as follows: in its $A_{N-1}$ integral, set the antisymmetric parameter to $a$, take the first $N$ parameters to be $t_j$, and take the remaining three parameters to be $s_1/a,s_2/a,s_3/a$. The theorem's auxiliary parameter is then $a^{N-2}Ts_1s_2s_3=pq/s_4$, and its inverse gamma factor $1/\Gamma(pq\,z_i^{-1}/s_4;p,q)$ equals $\Gamma(s_4z_i;p,q)$ by reflection.

\subsection{The $p\to0$ limit and the resulting integrals}

We retain $a$, all $s_I$, and $t_1,\ldots,t_{N-1}$ as fixed parameters, and set
\begin{equation}
 T_0=\prod_{j=1}^{N-1}t_j,\qquad
 C=a^{N-2}ST_0,\qquad t_N=\frac{pq}{C}.
 \label{eq:odd-antisymmetric-scaling}
\end{equation}
This choice satisfies the balancing condition. The fixed chirals have $R_0=0$ and give Neumann half-index factors. The last antifundamental has $R_0=2$ and contributes with Dirichlet boundary condition. The
reflection relation gives
\begin{equation}
\begin{gathered}
 \Gamma\!\left(\frac{pq}{Cz_i};p,q\right)
 =\frac{1}{\Gamma(Cz_i;p,q)},
 \qquad
 \Gamma\!\left(\frac{pq}{C} s_I;p,q\right)
 =\frac{1}{\Gamma(C/s_I;p,q)},
 \\
 \Gamma\!\left(\frac{pq}{C} at_j;p,q\right)
 =\frac{1}{\Gamma(C/(at_j);p,q)} .
\end{gathered}
 \label{eq:odd-antisymmetric-reflection}
\end{equation}
Moreover $\Gamma(T;p,q)=1/\Gamma(a^{N-2}S;p,q)$, and $a^nT=pq/(a^{n-2}S)$ for $N=2n$ while $a^nTs_I=pq/(a^{n-1}S/s_I)$ for $N=2n+1$. Taking $p\to0$ gives the electric half-index
\begin{equation}
\begin{aligned}
 \mathbb I_E={}&\frac{(q;q)_\infty^{N-1}}{N!}
 \int_{\mathbb T^{N-1}}
 \prod_{i=1}^{N-1}\frac{dz_i}{2\pi i z_i}
 \\
 &\times
 \frac{\prod_{\substack{1\leq i,j\leq N\\i\ne j}}(z_i/z_j;q)_\infty
 \prod_{i=1}^{N}(Cz_i;q)_\infty}
 {\prod_{i<j}(az_iz_j;q)_\infty
 \prod_{i=1}^{N}\prod_{I=1}^{4}(s_Iz_i;q)_\infty
 \prod_{i=1}^{N}\prod_{j=1}^{N-1}(t_j/z_i;q)_\infty}
\end{aligned}
\label{eq:odd-antisymmetric-half-electric}
\end{equation}
for either parity. For $N=2n+1$ the confined half-index becomes
\begin{equation}
\begin{aligned}
 \mathbb I_C^{(2n+1)}={}&
 \frac{(a^{N-2}S;q)_\infty
 \prod_{I=1}^{4}(C/s_I;q)_\infty
 \prod_{j=1}^{N-1}(C/(at_j);q)_\infty}
 {\prod_{I=1}^{4}(a^n s_I;q)_\infty
 \prod_{I=1}^{4}(a^{n-1}S/s_I;q)_\infty}
 \\
 &\times
 \frac{1}{\prod_{I=1}^{4}\prod_{j=1}^{N-1}
 (s_It_j;q)_\infty
 \prod_{1\leq j<k\leq N-1}(at_jt_k;q)_\infty},
\end{aligned}
\label{eq:odd-antisymmetric-half-confined}
\end{equation}
and for $N=2n$
\begin{equation}
\begin{aligned}
 \mathbb I_C^{(2n)}={}&
 \frac{(a^{N-2}S;q)_\infty
 \prod_{I=1}^{4}(C/s_I;q)_\infty
 \prod_{j=1}^{N-1}(C/(at_j);q)_\infty}
 {(a^{n};q)_\infty\,(a^{n-2}S;q)_\infty
 \prod_{1\le I<J\le4}(a^{n-1}s_Is_J;q)_\infty}
 \\
 &\times
 \frac{1}{\prod_{I=1}^{4}\prod_{j=1}^{N-1}
 (s_It_j;q)_\infty
 \prod_{1\leq j<k\leq N-1}(at_jt_k;q)_\infty}.
\end{aligned}
\label{eq:even-antisymmetric-half-confined}
\end{equation}
The resulting half-index identities are
\begin{equation}
 \mathbb I_E=\mathbb I_C^{(2n+1)}\quad (N=2n+1),\qquad
 \mathbb I_E=\mathbb I_C^{(2n)}\quad (N=2n).
 \label{eq:odd-antisymmetric-half-equality}
\end{equation}

The limit can be taken inside the contour integral in a nonempty convergence region, as discussed in appendix~\ref{app:convergence}. The resulting identities extend to other parameter domains by analytic continuation wherever valid.
\subsection{Half-index identities and boundary dualities}

The limiting gauge content consists of a Neumann vector, a Neumann antisymmetric chiral, four Neumann fundamentals, $N-1$ Neumann antifundamentals and one Dirichlet antifundamental. Their boundary gauge anomaly vanishes,
\begin{equation}
 \mathcal A_{\mathrm{gauge}}
 =N-\frac{N-2}{2}-4\left(\frac12\right)
   -(N-1)\left(\frac12\right)+\frac12=0.
 \label{eq:odd-antisymmetric-anomaly}
\end{equation}
The field content of the dual pairs for even and odd $N$ is given in Table~\ref{tab:antisym}. 

Under the identification
\begin{equation}
\begin{gathered}
 a\to q^{r_A/2}A,\qquad s_I\to q^{r_a/2}a\,x_I,\\
 t_\alpha\to q^{r_b/2}b\,\tilde x_\alpha\quad(\alpha\le N-1),\qquad z_i\to s_i ,
\end{gathered}
\end{equation}
where on the right-hand sides $A,a,b$ denote the fugacities of
\cite{OkazakiSmith2023}, our electric and confined expressions reproduce
the half-indices proposed there.\footnote{For even $N=2n$, two
typographical corrections are needed in the confined half-index of
\cite{OkazakiSmith2023}: the denominator factor
$(q^{((N-2)r_A+4r_a)/2}A^{N-2}a^4;q)_\infty$
should read
$(q^{((n-2)r_A+4r_a)/2}A^{n-2}a^4;q)_\infty$,
and $4r_b$ in the exponent of the first numerator factor should read
$4r_a$. The corrected expression agrees with
eq.~\eqref{eq:even-antisymmetric-half-confined} and with our numerical
checks.}
These identities were originally proposed on the basis of conjectural
boundary confining dualities and supported by anomaly matching and a
$q$-series check for $N=5$. For $N\leq4$, they reduce to previously known
cases, while $N=5$ and $N=6$ provide the first new examples for odd and
even $N$, respectively.

We establish these identities analytically for all $N\geq5$, for both even and odd $N$. The proof follows from the known four-dimensional elliptic hypergeometric integral evaluations and the $p\to0$ limit derived above.\footnote{As a numerical consistency check, we evaluated both sides of the half-index identities for $N=5$ and $N=6$ at $q=0.3$, using periodic trapezoidal quadrature for the multidimensional integrals. For representative complex fugacities, the electric and confined expressions agree with relative differences of approximately $10^{-10}$ and $10^{-6}$, respectively. For real fugacities, the corresponding relative differences are approximately $10^{-12}$ and $10^{-9}$.}

\begin{table}[ht]
\centering
\small
\renewcommand{\arraystretch}{1.3}
\begin{tabular}{|c|c|c|c|c|}
\hline
 & $SU(N)$ & $SU(4)\times SU(N-1)$ & fugacity & b.c.\\
\hline
\multicolumn{5}{|l|}{\textbf{Theory A}}\\
\hline
VM & adj & $(1,1)$ & - & $\mathcal N$\\
$\Phi$ & $\Lambda^2$ & $(1,1)$ & $a$ & $\mathcal N$\\
$Q_I$ & $\square$ & $(\square,1)$ & $s_I$ & $\mathcal N$\\
$\tilde Q_j$ & $\overline{\square}$ & $(1,\square)$ & $t_j$ & $\mathcal N$\\
$\tilde Q_N$ & $\overline{\square}$ & $(1,1)$ & $pq/C$ & $\mathcal D$\\
\hline
\multicolumn{5}{|l|}{\textbf{Theory B, both parities}}\\
\hline
$M_{Ij}\sim Q_I\tilde Q_j$ & $1$ & $(\square,\square)$ & $s_It_j$ & $\mathcal N$\\
$M_{jk}\sim \Phi\tilde Q_j\tilde Q_k$ & $1$ & $(1,\Lambda^2)$ & $at_jt_k$ & $\mathcal N$\\
$\tilde M_I\sim Q_I\tilde Q_N$ & $1$ & $(\square,1)$ & $pq/(C/s_I)$ & $\mathcal D$\\
$\tilde M_j\sim \Phi\tilde Q_j\tilde Q_N$ & $1$ & $(1,\square)$ & $pq/(C/(at_j))$ & $\mathcal D$\\
$\tilde B\sim \epsilon\,\tilde Q_1\cdots\tilde Q_N$ & $1$ & $(1,1)$ & $pq/(a^{N-2}S)$ & $\mathcal D$\\
\hline
\multicolumn{5}{|l|}{\textbf{Theory B, $N=2n+1$}}\\
\hline
$B_I\sim \epsilon\,\Phi^nQ_I$ & $1$ & $(\square,1)$ & $a^ns_I$ & $\mathcal N$\\
$\bar B^I\sim \epsilon\,\Phi^{n-1}Q_JQ_KQ_L$ & $1$ & $(\overline{\square},1)$ & $a^{n-1}S/s_I$ & $\mathcal N$\\
\hline
\multicolumn{5}{|l|}{\textbf{Theory B, $N=2n$}}\\
\hline
$\phi\sim \epsilon\,\Phi^n$ & $1$ & $(1,1)$ & $a^n$ & $\mathcal N$\\
$B_{IJ}\sim \epsilon\,\Phi^{n-1}Q_IQ_J$ & $1$ & $(\Lambda^2,1)$ & $a^{n-1}s_Is_J$ & $\mathcal N$\\
$B\sim \epsilon\,\Phi^{n-2}Q_1Q_2Q_3Q_4$ & $1$ & $(1,1)$ & $a^{n-2}S$ & $\mathcal N$\\
\hline
\end{tabular}
\caption{Dual pairs for $SU(N)$ with an antisymmetric chiral, four fundamentals and $N$ antifundamentals, with $I=1,\ldots,4$, $j,k=1,\ldots,N-1$, $S=\prod_Is_I$ and $C=a^{N-2}S\prod_{j=1}^{N-1}t_j$. In the row of $\bar B^I$, $\{I,J,K,L\}=\{1,2,3,4\}$.}
\label{tab:antisym}
\end{table}

\subsection{Comparison with the Gustafson-Rakha integral}

Taking $p\to0$ in Theorem~7 of \cite{Spiridonov2003} at fixed parameters, so that the factor $1/\Gamma(Az_j^{-1};p,q)$ with the auxiliary parameter $A=a^{N-2}Ts_1s_2s_3$ tends to $(Az_j^{-1};q)_\infty$ (equivalently, scaling the fourth fundamental $s_4=pq/A$ as $p$), gives the Gustafson-Rakha integral \cite{GustafsonRakha2000}, in which the antisymmetric tensor is accompanied by three Neumann fundamentals, $N$ Neumann antifundamentals and a numerator factor of fundamental type. This is the degeneration identified in \cite{SpiridonovVartanov2011,Spiridonov2008,Rosengren:2017ujl} and corresponds to the boundary confining duality of \cite{OkazakiSmith2023}; the associated residue sums are the Gustafson-Rakha summations and their elliptic extension \cite{GustafsonRakha2000,Rosengren2017}. The scaling \eqref{eq:odd-antisymmetric-scaling}, with an antifundamental scaled as $p$, instead leaves four Neumann fundamentals, $N-1$ Neumann antifundamentals and a numerator factor of antifundamental type, and yields the different identities \eqref{eq:odd-antisymmetric-half-equality}. To our knowledge, these $q$-beta integral evaluations do not appear in \cite{GustafsonRakha2000,Schlosser2020,Rosengren:2017ujl,Rosengren2017,Spiridonov2008}.

\section{Conclusions}
We have considered the $p\to0$ limit of four-dimensional $\mathcal N=1$ superconformal indices and obtained the corresponding half-index identities for three-dimensional $\mathcal N=2$ theories. The limit is controlled by the scaling of chiral multiplet fugacities. A multiplet with $R_0=0$ gives a Neumann contribution, one with $R_0=2$ gives a Dirichlet contribution, and one with $0<R_0<2$ does not contribute. The vector multiplet gives a Neumann contribution, and the four-dimensional gauge-R anomaly cancellation becomes the boundary gauge anomaly cancellation. This gives a superconformal-index interpretation of the $q$-hypergeometric functions that arise as $p\to0$ limits of elliptic hypergeometric integrals, a question raised in \cite{SpiridonovVartanov2011}.

For $SU(2)$, $USp(2N)$, $G_2$, and $SU(N)$ with $N+1$ flavors, the limiting identities are known $q$-beta integral evaluations of Nassrallah-Rahman and Gustafson, and they reproduce the half-index identities of the boundary confining dualities of \cite{OkazakiSmith2023,OkazakiSmithExceptional2023}. For $SU(N)$ with an antisymmetric tensor, four fundamentals and $N$ antifundamentals, we provide an analytic proof of the half-index identities of \cite{OkazakiSmith2023} for all $N\ge5$, including both even and odd $N$. To our knowledge, the resulting $q$-beta integral evaluations have not previously appeared in the literature and are distinct from the Gustafson-Rakha integral \cite{GustafsonRakha2000}. Other scalings of the same four-dimensional identities give the Askey-Wilson integral and Gustafson's Askey-Wilson-type integrals (appendix~\ref{app:alternative}).

We emphasise that this correspondence holds at the level of protected indices. It would be interesting to investigate whether it can be derived directly from supersymmetric localisation on a suitable degenerating geometry and to determine under which conditions it extends to a relation between full theories with boundary. Applying residue calculus to the new integrals of section~\ref{sec:antisym}, as was done for the Gustafson-Rakha integral \cite{GustafsonRakha2000,Rosengren2017}, may lead to new multivariable summation formulae.

Another direction concerns line defects, whose duals for the Nassrallah-Rahman and Gustafson-Rakha cases were left open in \cite{OkazakiSmith2026}. A suitable insertion of theta functions in the elliptic integrand reduces to a Wilson-line insertion in the $p\to0$ limit, which suggests a four-dimensional surface-defect origin for these line operators that we leave for future work.

\acknowledgments
We would like to thank Ali Mert Yetkin, who participated in the initial stages of this work, for valuable discussions. Both authors are supported by the Istanbul Integrability and Stringy Topics Initiative (\href{https://istringy.org/}{istringy.org}).

\appendix
\section{Elliptic gamma function and its $p\to0$ limits}
\label{app:gamma}

We use the elliptic gamma function \cite{Spiridonov2003,Spiridonov2008}
\begin{equation}\label{eq:egamma}
\Gamma(t; p, q) = \prod_{i,j=0}^{\infty} \frac{1 - t^{-1} p^{i+1} q^{j+1}}{1 - t p^i q^j},\qquad |p|,|q|<1 ,
\end{equation}
which, for $|pq|<|t|<1$, can also be written as
\[
\exp\!\left[\sum_{n\geq 1}\frac{1}{n}\frac{t^n-p^nq^nt^{-n}}{(1-p^n)(1-q^n)}\right].
\]
It satisfies the reflection formula $\Gamma(t;p,q)\,\Gamma(pq/t;p,q)=1$.

Let $t=p^\alpha c$ with $c\ne0$ fixed. For $\alpha=0$, all numerator factors of~\eqref{eq:egamma} tend to $1$, while the denominator factors with $i=0$ survive:
\begin{equation}
\lim_{p\to0}\Gamma(c;p,q)=\prod_{j=0}^{\infty}\frac{1}{1-cq^j}=\frac{1}{(c;q)_\infty}.
\end{equation}
For $\alpha=1$, the product becomes $\prod_{i,j}(1-c^{-1}p^iq^{j+1})/(1-cp^{i+1}q^j)$; now the denominator tends to $1$ and the numerator factors with $i=0$ survive:
\begin{equation}
\lim_{p\to0}\Gamma(pc;p,q)=\prod_{j=0}^{\infty}(1-c^{-1}q^{j+1})=(q/c;q)_\infty .
\end{equation}
For $0<\alpha<1$, every factor in the numerator and in the denominator carries a positive power of $p$, so the limit is $1$. These are the three cases of eq.~\eqref{eq:limits}.

\subsection{Justification of the limit}
\label{app:convergence}
We now justify taking the limit in a nonempty convergence region. Take $|q|<1$ and fixed nonzero parameters with $|a|,|s_I|,|t_j|<1$ for $j<N$, $|C|<1$, and $|s_I/a|<1$ for $I=1,2,3$, as required by the parametrisation of \cite{Spiridonov2003}. For sufficiently small $|p|$, the unit torus remains an admissible contour. After applying the reflection relations \eqref{eq:odd-antisymmetric-reflection}, the infinite-product representation of the elliptic gamma function gives uniform convergence of the integrand on the unit torus, while the conditions on the fugacities ensure that the denominator factors remain uniformly bounded away from zero. Dominated convergence therefore allows the limit $p\to0$ to be taken inside the integral. The identity can be extended to other parameter domains by analytic continuation wherever valid.
\section{Anomalies and superpotentials}
\label{sec:anomaly}

The balancing conditions of the elliptic integrals reflect the cancellation of the four-dimensional gauge-R anomaly,
\begin{equation}\label{eq:anomaly}
\mathrm{Tr}\,R_0\,G\,G=T(\mathrm{adj})+\sum_{\text{chirals}}\big(R_0-1\big)\,T(\rho)=0 .
\end{equation}
Chiral multiplets with $R_0=0$ and $R_0=2$ contribute $-T(\rho)$ and $+T(\rho)$. Fields with $R_0=1$ do not contribute, and a pair with $R_0=2\alpha$ and $2(1-\alpha)$ in the same representation contributes zero in total. When only $R_0=0$ and $R_0=2$ fields survive,
\begin{equation}
T(\mathrm{adj})-\sum_{\text{Neumann}}T(\rho)+\sum_{\text{Dirichlet}}T(\rho)=0 ,
\end{equation}
which is the boundary gauge anomaly cancellation condition for a three-dimensional theory with Neumann vector multiplet \cite{DimofteGaiottoPaquette2018,OkazakiSmith2021,OkazakiSmith2023}. The superpotential must satisfy $R_0(W)=2$. If it contains a Dirichlet field, its boundary restriction vanishes upon imposing the Dirichlet condition \cite{DimofteGaiottoPaquette2018}.

\section{Alternative scaling limits}
\label{app:alternative}

The same four-dimensional identities admit other scaling limits. In the three examples below, two chiral multiplets acquire $R_0=2\alpha$ and $2(1-\alpha)$ with $0<\alpha<1$ and decouple individually, while the meson built from them survives as a Dirichlet singlet. In each case the resulting $q$-beta integral is known, and the boundary anomaly cancels as discussed in appendix~\ref{sec:anomaly}.

\subsection{$SU(2)$: the Askey-Wilson integral}

In the elliptic beta integral~\eqref{eq:ebi} we keep $t_i=a_i$ ($i\le4$) fixed and set $t_5=p^{1/2}c_5$, $t_6=p^{1/2}c_6$ with $c_5c_6=q/(a_1a_2a_3a_4)$. The meson $M_{56}$ has fugacity $t_5t_6=pq/(a_1a_2a_3a_4)$ and gives the Dirichlet factor $(a_1a_2a_3a_4;q)_\infty$, while $M_{i5}$ and $M_{i6}$ decouple. Taking the limit gives the Askey-Wilson integral,
\begin{equation}
\frac{(q;q)_\infty}{2}
\oint_{\mathbb{T}}
\frac{(z^{\pm2};q)_\infty}
{\prod_{i=1}^{4}(a_i z^{\pm1};q)_\infty}
\frac{dz}{2\pi iz}
=
\frac{(a_1a_2a_3a_4;q)_\infty}
{\prod_{1\leq i<j\leq4}(a_i a_j;q)_\infty},
\label{eq:askey-wilson-limit}
\end{equation}
which corresponds to the boundary confining duality of \cite{OkazakiSmith2023}; see Table~\ref{tab:AW}. Thus the Nassrallah-Rahman and Askey-Wilson integrals arise from different scalings of the same index.

\begin{table}[ht]
\centering
\small
\renewcommand{\arraystretch}{1.3}
\begin{tabular}{|c|c|c|c|c|}
\hline
 & $SU(2)$ & $SU(4)$ & fugacity & b.c.\\
\hline
\multicolumn{5}{|l|}{\textbf{Theory A}}\\
\hline
VM & adj & $1$ & - & $\mathcal N$\\
$Q_i$ & $\square$ & $\square$ & $a_i$ & $\mathcal N$\\
$Q_5,\ Q_6$ & $\square$ & $1$ & $p^{1/2}c_5,\ p^{1/2}c_6$ & -\\
\hline
\multicolumn{5}{|l|}{\textbf{Theory B}}\\
\hline
$M_{ij}\sim Q_iQ_j$ & $1$ & $\Lambda^2$ & $a_ia_j$ & $\mathcal N$\\
$M_{i5},\ M_{i6}$ & $1$ & $\square$ & $\propto p^{1/2}$ & -\\
$M_{56}\sim Q_5Q_6$ & $1$ & $1$ & $pq/(a_1a_2a_3a_4)$ & $\mathcal D$\\
\hline
\end{tabular}
\caption{Dual pair for the Askey-Wilson limit of the $SU(2)$ theory with six doublets, $i,j=1,\ldots,4$.}
\label{tab:AW}
\end{table}

\subsection{$USp(2N)$: Gustafson's $C_N$ Askey-Wilson-type integral}

In the $USp(2N)$ identity of section~\ref{sec:USp} we keep $t_1,\ldots,t_{2N+2}$ fixed and set
\begin{equation}
t_{2N+3}=p^\alpha c,\qquad
 t_{2N+4}=p^{1-\alpha}\frac{q}{Ac},\qquad
A=\prod_{m=1}^{2N+2}t_m .
\label{eq:usp-alt-scaling}
\end{equation}
The meson $Q_{2N+3}Q_{2N+4}$ has fugacity $pq/A$ and gives the Dirichlet factor $(A;q)_\infty$. The limit is Gustafson's $C_N$ Askey-Wilson-type integral \cite{Gustafson1994},
\begin{multline}
\frac{(q;q)_\infty^N}{2^NN!}
\int_{\mathbb{T}^{N}}
\prod_{1\le i<j\le N}
(z_i^{\pm1}z_j^{\pm1};q)_\infty
\prod_{j=1}^{N}
\frac{(z_j^{\pm2};q)_\infty}
{\prod_{m=1}^{2N+2}
(t_mz_j^{\pm1};q)_\infty}
\frac{dz_j}{2\pi i z_j}\\
=\frac{(A;q)_\infty}
{\prod_{1\le m<n\le 2N+2}
(t_mt_n;q)_\infty},
\label{eq:usp-alt}
\end{multline}
corresponding to the boundary confining duality of \cite{OkazakiSmith2023}; see Table~\ref{tab:USpalt}.

\begin{table}[ht]
\centering
\small
\renewcommand{\arraystretch}{1.3}
\begin{tabular}{|c|c|c|c|c|}
\hline
 & $USp(2N)$ & $SU(2N+2)$ & fugacity & b.c.\\
\hline
\multicolumn{5}{|l|}{\textbf{Theory A}}\\
\hline
VM & adj & $1$ & - & $\mathcal N$\\
$Q_m$ & $\square$ & $\square$ & $t_m$ & $\mathcal N$\\
$Q_{2N+3}$ & $\square$ & $1$ & $p^\alpha c$ & -\\
$Q_{2N+4}$ & $\square$ & $1$ & $p^{1-\alpha}q/(Ac)$ & -\\
\hline
\multicolumn{5}{|l|}{\textbf{Theory B}}\\
\hline
$M_{mn}\sim Q_mQ_n$ & $1$ & $\Lambda^2$ & $t_mt_n$ & $\mathcal N$\\
$M_{m,2N+3},\ M_{m,2N+4}$ & $1$ & $\square$ & $\propto p^{\alpha},\,p^{1-\alpha}$ & -\\
$Q_{2N+3}Q_{2N+4}$ & $1$ & $1$ & $pq/A$ & $\mathcal D$\\
\hline
\end{tabular}
\caption{Dual pair for the alternative scaling of the $USp(2N)$ theory, $m,n=1,\ldots,2N+2$, $A=\prod_{m=1}^{2N+2}t_m$ and $0<\alpha<1$.}
\label{tab:USpalt}
\end{table}

\subsection{$SU(N)$: Gustafson's $A_{N-1}$ Askey-Wilson-type integral}

In the $SU(N)$ identity of section~\ref{sec:SUN} we keep $s_i$ and $t_i$, $i\le N$, fixed and set
\begin{equation}
S_0=\prod_{i=1}^{N}s_i,\qquad
T_0=\prod_{j=1}^{N}t_j,\qquad
s_{N+1}=p^\alpha c,\qquad
t_{N+1}=p^{1-\alpha}\frac{q}{S_0T_0c}.
\label{eq:su-alt-scaling}
\end{equation}
The meson $Q_{N+1}\tilde Q_{N+1}$ has fugacity $pq/(S_0T_0)$ and gives the Dirichlet factor $(S_0T_0;q)_\infty$, while the baryon $Q_1\cdots Q_N$ and antibaryon $\tilde Q_1\cdots\tilde Q_N$ remain Neumann. The limit is Gustafson's $A_{N-1}$ Askey-Wilson-type integral \cite{Gustafson1994},
\begin{multline}
\frac{(q;q)_\infty^{N-1}}{N!}
\int_{\mathbb{T}^{N-1}}
\frac{\prod_{i\ne j}(z_i/z_j;q)_\infty}{\prod_{i,j=1}^{N}
(s_i z_j;q)_\infty(t_i z_j^{-1};q)_\infty}
\prod_{j=1}^{N-1}\frac{dz_j}{2\pi i z_j}\\
=
\frac{(S_0T_0;q)_\infty}
{(S_0;q)_\infty(T_0;q)_\infty
\prod_{i,j=1}^{N}(s_it_j;q)_\infty},
\label{eq:su-alt}
\end{multline}
corresponding to the boundary confining duality of \cite{OkazakiSmith2023}; see Table~\ref{tab:SUNalt}.

\begin{table}[ht]
\centering
\small
\renewcommand{\arraystretch}{1.3}
\begin{tabular}{|c|c|c|c|c|}
\hline
 & $SU(N)$ & $SU(N)\times SU(N)$ & fugacity & b.c.\\
\hline
\multicolumn{5}{|l|}{\textbf{Theory A}}\\
\hline
VM & adj & $(1,1)$ & - & $\mathcal N$\\
$Q_i$ & $\square$ & $(\square,1)$ & $s_i$ & $\mathcal N$\\
$\tilde Q_j$ & $\overline{\square}$ & $(1,\square)$ & $t_j$ & $\mathcal N$\\
$Q_{N+1}$ & $\square$ & $(1,1)$ & $p^\alpha c$ & -\\
$\tilde Q_{N+1}$ & $\overline{\square}$ & $(1,1)$ & $p^{1-\alpha}q/(S_0T_0c)$ & -\\
\hline
\multicolumn{5}{|l|}{\textbf{Theory B}}\\
\hline
$M_{ij}\sim Q_i\tilde Q_j$ & $1$ & $(\square,\square)$ & $s_it_j$ & $\mathcal N$\\
$B\sim \epsilon\,Q_1\cdots Q_N$ & $1$ & $(1,1)$ & $S_0$ & $\mathcal N$\\
$\tilde B\sim \epsilon\,\tilde Q_1\cdots\tilde Q_N$ & $1$ & $(1,1)$ & $T_0$ & $\mathcal N$\\
$Q_{N+1}\tilde Q_{N+1}$ & $1$ & $(1,1)$ & $pq/(S_0T_0)$ & $\mathcal D$\\
others & $1$ & various & $\propto p^{\alpha},\,p^{1-\alpha}$ & -\\
\hline
\end{tabular}
\caption{Dual pair for the alternative scaling of the $SU(N)$ theory with $N+1$ flavors, $i,j=1,\ldots,N$, $S_0=\prod_{i=1}^Ns_i$, $T_0=\prod_{j=1}^Nt_j$ and $0<\alpha<1$. The entry ``others'' denotes the decoupled composites $M_{i,N+1}$, $M_{N+1,j}$ and the baryons and antibaryons containing $Q_{N+1}$ or $\tilde Q_{N+1}$.}
\label{tab:SUNalt}
\end{table}

\bibliographystyle{utphys}
\bibliography{references}

\end{document}